\documentclass[table]{article}
\usepackage{listings}
\usepackage{graphicx}
\usepackage{booktabs}
\usepackage{amsmath}
\usepackage{float}
\usepackage{hyperref}
\usepackage{setspace}
\usepackage[top=1.5in, bottom=1.5in, left= 1.0in, right=1.0in]{geometry}
\usepackage{xcolor}
\usepackage{subcaption}
\usepackage[round]{natbib}

\usepackage[utf8]{inputenc}
\usepackage[T1]{fontenc}
\usepackage{listingsutf8}

\usepackage{inconsolata}
\usepackage[most]{tcolorbox} 

\lstdefinestyle{monoblock}{
basicstyle=\ttfamily\normalsize,
breaklines=true,
breakatwhitespace=false,
columns=fullflexible,
keepspaces=true,
frame=single,
rulecolor=\color{black!15},
xleftmargin=0pt,
xrightmargin=0pt,
}

\lstnewenvironment{Monoblock}[1][]{
\lstset{style=monoblock,#1}
}{}

\title{
\vspace{-6em}
\includegraphics[width=1\linewidth]{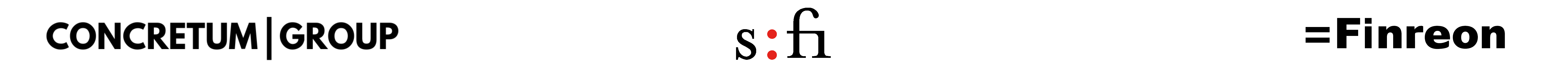}
\\[3em]
\huge{ChatGPT in Systematic Investing}\\[0em]
\Large{Enhancing Risk-Adjusted Returns with LLMs}
}
\vspace{-3em}
\author{
        Nikolas Anic, Andrea Barbon, Ralf Seiz, Carlo Zarattini\thanks{\fontsize{8}{8}\selectfont 
          Nikolas Anic, (\href{mailto:nikolas.anic@finreon.ch}{nikolas.anic@finreon.ch}) Finreon AG, Oberer Graben 3, CH-9000, St. Gallen.
          Andrea Barbon (\href{mailto:andrea.barbon@unisg.ch}{andrea.barbon@unisg.ch}), University of St.Gallen, Dufourstrasse 50, 9000 St. Gallen, Switzerland.
          Ralf Seiz (\href{mailto:ralf.seiz@finreon.ch}{ralf.seiz@finreon.ch}) Finreon AG and University of St.Gallen, Oberer Graben 3, CH-9000, St. Gallen.
          Carlo Zarattini (\href{mailto:carlo@concretumgroup.com}{carlo@concretumgroup.com}), Concretum Group, Viale Carlo Cattaneo 1, 6900 Lugano, Switzerland.
          Andrea Barbon is the corresponding author.
          }
}

\date{}

\begin{document}

\setstretch{1.5}
\setlength{\parindent}{0pt}
\setlength{\parskip}{1.5em}%
\renewcommand{\footnotesize}{\fontsize{8pt}{11pt}\selectfont}

\maketitle

\begin{abstract}
    This paper investigates whether large language models (LLMs) can improve cross-sectional momentum strategies by extracting predictive signals from firm-specific news. 
    We combine daily U.S. equity returns for S\&P 500 constituents with high-frequency news data and use prompt-engineered queries to ChatGPT that inform the model when a stock is about to enter a momentum portfolio. 
    The LLM evaluates whether recent news supports a continuation of past returns, producing scores that condition both stock selection and portfolio weights. 
    An LLM-enhanced momentum strategy outperforms a standard long-only momentum benchmark, delivering higher Sharpe and Sortino ratios both in-sample and in a truly out-of-sample period after the model's pre-training cut-off. 
    These gains are robust to transaction costs, prompt design, and portfolio constraints, and are strongest for concentrated, high-conviction portfolios. 
    The results suggest that LLMs can serve as effective real-time interpreters of financial news, adding incremental value to established factor-based investment strategies.
\end{abstract}
\newpage

\section{Introduction}

    Large Language Models (LLMs) have demonstrated utility across a wide range of domains, from legal document analysis to medical diagnostics and customer service automation. In the domain of financial wealth management, however, empirical evidence on the effectiveness of LLMs remains scarce.
    Many studies have linked the existence of factor anomalies, particularly momentum, to the release of firm-specific fundamental news and the gradual incorporation of such information into prices. 
    The literature suggests that investors underreact to news in the short run, leading to predictable price drifts that can be exploited through momentum strategies. 
    For instance, post-earnings announcement drift and other delayed reactions to public disclosures highlight how information processing frictions create opportunities for return predictability\citep{bernard1989post}. 
    Building on this intuition, we posit that the ability of LLMs to interpret and synthesize textual information in real time can be used to identify news that is likely to trigger price adjustments. 
    In turn, this should allow to better predict if the observed positive past returns are likely to be driven by momentum and continue in the near future. 
    This leads to the conjecture that combining traditional momentum signals with LLM-driven insights extracted from stock-specific news can lead to more informed and potentially better-performing momentum portfolios.

    Hence, in this paper we ask whether LLMs can be employed to enhance factor-based trading strategies, more specifically momentum strategies, by extracting signals from stock-specific news. 
    To empirically test our conjecture, we assemble a novel dataset combining several high-frequency information sources. First, we utilize a historical dataset of U.S. equity returns at daily frequency, allowing us to compute standard cross-sectional momentum signals and evaluate the performance of trading strategies over time. This dataset serves as the foundation for measuring baseline momentum effects and benchmarking any improvement attributable to the incorporation of LLM-based signals.

    Second, we obtain a granular database of stock-specific news articles with minute-level timestamps from the Stock News API. This fine temporal resolution is essential for aligning news content with market reactions and isolating the immediate informational impact of news events. By focusing on firm-level disclosures and developments, rather than broader macroeconomic headlines, we aim to capture the specific narratives that could plausibly drive momentum at the individual stock level.

    Third, we apply prompt engineering techniques to interact with state-of-the-art ChatGPT models developed by OpenAI. These prompts are carefully designed to extract semantic signals from the text of each news article, such as its relevance to the firm, and whether the news is likely to reinforce or contradict existing price trends. Importantly, the prompts explicitly inform the LLM that the stock under consideration is about to enter a baseline momentum portfolio, thereby framing the analysis in the context of a potential trading decision. This allows the model to assess the news in light of the expected continuation of past returns, enhancing its ability to detect signals that either support or challenge the momentum hypothesis. This LLM-driven layer of interpretation enables us to construct a dynamic signal that augments the traditional momentum indicator with insights extracted from contemporaneous textual data.

    Fourth, we focus on a trading strategy based on a simple yet well-established technical signal: cross-sectional momentum. Stocks are ranked based on their past returns over a fixed lookback window, and positions are taken accordingly. We then enrich this strategy by conditioning portfolio weights on the LLM-generated signals from firm-specific news. This design allows us to directly assess whether real-time interpretation of news content by a large language model can improve the informational efficiency and performance of momentum-based strategies.

    Our empirical findings provide support for the value of incorporating LLM-based insights into momentum strategies. The LLM-enhanced momentum portfolio consistently outperforms the baseline strategy, both in-sample and out-of-sample, delivering economically meaningful improvements in risk-adjusted returns. These performance gains remain robust even after accounting for conservative estimates of transaction costs of 2 bps.

    We find that a monthly rebalancing frequency yields significantly better results than a weekly one. This is likely driven by lower turnover and reduced transaction costs, which disproportionately affect higher-frequency strategies. Moreover, the chronological inconsistency of the employed LLM, due to its pre-training cutoff, does not appear to inflate performance. On the contrary, results are even stronger when we restrict attention to the period after the model's training data ends. This suggests that the model's ability to interpret news relies on its general language understanding rather than on memorized historical content.

    We also compare alternative prompt formulations and find that simpler prompts lead to slightly better performance, although the difference is not statistically significant. Importantly, applying diversification constraints to portfolio weights reduces the Sharpe ratio slightly, but results in substantial improvements in terms of risk-management. This indicates that the raw LLM-driven signals tend to produce overly concentrated positions, which can be mitigated through appropriate portfolio construction techniques.

    Finally, we document a key difference in how performance varies with portfolio size. We understand that the LLM-enhanced strategy performs better when applied to a smaller set of high-conviction signals. This suggests that the LLM’s informational edge is strongest in identifying particularly news-sensitive opportunities rather than offering broad-based cross-sectional coverage.

\subsection{Related Literature}

    This paper builds on three main strands of the literature: the extensive research on momentum and market underreaction, the growing body of work on textual analysis in finance, and the emerging application of large language models (LLMs) in asset pricing and investment strategies.

    A foundational literature in empirical asset pricing documents the existence of momentum profits, whereby stocks with strong recent returns tend to outperform in the near future \citep{jegadeesh1993returns, chan1996momentum}. Several studies attribute these return patterns to delayed price adjustments following firm-specific information events, suggesting that investors underreact to news in the short term \citep{bernard1989post, hong1999unified}. This underreaction gives rise to return predictability, particularly around earnings announcements and other public disclosures. Building on this, cross-sectional momentum strategies exploit price trends across stocks, often assuming that information is gradually incorporated into prices over time.

    A parallel literature has explored the use of textual data as a source of predictive signals for financial markets. Early contributions examine the sentiment content of news articles, earnings reports, and press releases to explain asset price movements \citep{tetlock2007giving, loughran2011liability}. Advances in natural language processing (NLP) have enabled more sophisticated extraction of sentiment and topic-based signals from unstructured text \citep{nassirtoussi2014text, ke2019predicting}. Recent studies have integrated news analytics into asset pricing models, showing that text-based signals and machine learning models can complement traditional return predictors \citep{gu2020empirical}.

    More recently, large language models such as GPT-3 and GPT-4 have opened new avenues for real-time textual analysis. While prior NLP approaches often relied on keyword dictionaries, sentiment scores, or supervised classifiers, LLMs provide a flexible, zero-shot interface for extracting nuanced information from unstructured text. A growing number of studies have explored the potential of LLMs in economic and financial applications, including forecasting, decision support, and sentiment extraction \citep{lopez2023can, kim2024financial}. However, few papers have examined whether LLMs can improve factor-based strategies in systematic investing. Notably, \citet{lopez2023can} show that ChatGPT can predict market reactions to news headlines, but applications to cross-sectional strategies remain limited.

    Our paper contributes to this emerging literature by evaluating whether prompt-engineered LLMs can enhance momentum investing by interpreting firm-specific news at scale. We are among the first to test the integration of LLM-generated insights into real-time portfolio construction, assessing the economic value of such signals in a cross-sectional asset pricing framework.

\section{Methodology}\label{sec:methodology}

    We start by replicating the classical cross-sectional momentum strategy outlined in \cite{jegadeesh1993returns} and \cite{carhart1997persistence}, sorting stocks each month based on their past 12-month returns (excluding the most recent month) and constructing a long-only portfolio by buying the top two deciles. We focus on the long leg to better reflect realistic investment constraints and institutional practices. Furthermore, we include the top two deciles to ensure performance comparability with the LLM portfolio construction.
    To ensure liquidity and focus on implementable portfolios, we restrict our universe to stocks that are constituents of the S\&P 500 index. This focus facilitates a cleaner implementation of the strategy, avoids biases due to illiquid small-cap stocks, and aligns with the investment universe typically considered by institutional asset managers.

    \begin{figure}[t]
        \centering\includegraphics[width=1\textwidth]{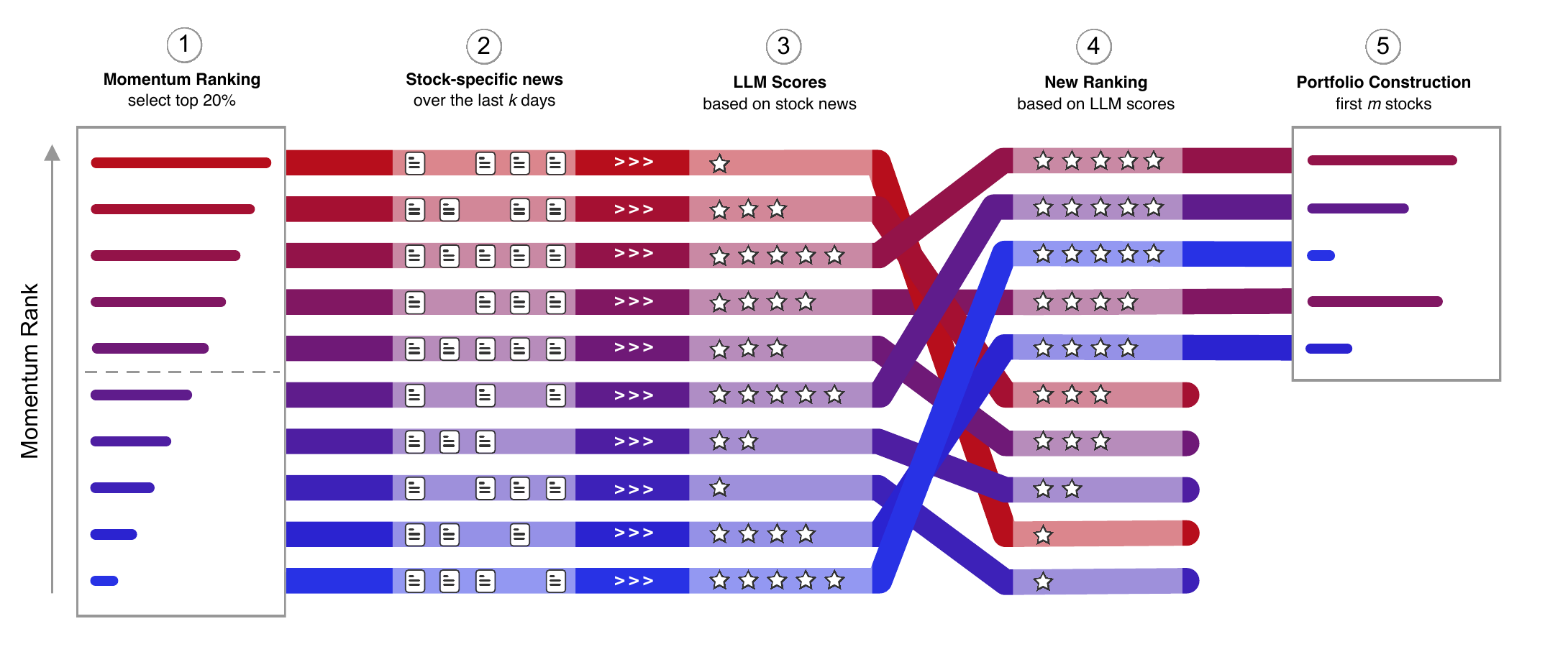}
        \caption{Flow diagram describing the process integrating news, the LLM signal, and portfolio construction.}
        \label{fig:flow}
    \end{figure}

    We then modify the composition of the momentum portfolio using signals generated by an LLM model. To create these signals, we feed the LLM with stock-specific news related to a broader set of momentum candidates, namely the top two deciles based on past returns, instead of just the top decile. The LLM is prompted to assess whether each stock should be included in or excluded from the final momentum portfolio. In addition, we allow the portfolio weights to be adjusted according to the strength of the LLM signal.
    The full methodology is detailed in the following subsections and visually summarized in Figure \ref{fig:flow}.

\subsection{Data Sources}

    Our empirical analysis combines three distinct datasets, each serving a specific role in constructing and evaluating the LLM-enhanced momentum strategy.

    First, we use an internal Finreon dataset, similar in structure and coverage to the CRSP database, which provides daily total returns and market capitalization data for U.S. equities. This dataset forms the backbone for computing baseline momentum signals, applying liquidity and index membership filters, and implementing portfolio construction.

    Second, we obtain high-frequency stock-specific news from the \emph{Stock News API}. This dataset includes publication timestamps with second-level precision, enabling precise alignment of news items with intraday market activity. For each news item, we record the title of the article, a short summary of its content, and the publication source. News is sourced from reputable outlets such as CNBC, Zacks, Bloomberg, The Motley Fool, Fox Business, and The Street, among others. The focus on firm-level news allows us to isolate informational events that are most likely to drive short-term price adjustments.

    Third, we retrieve the U.S. risk-free rate from the \emph{Federal Reserve Economic Data (FRED)} database. The risk-free rate series is used to compute excess returns for portfolio performance evaluation and to ensure consistency in risk-adjusted performance metrics.

\subsection{LLM-Enhanced Momentum Portfolio}

    At each month-end, we begin by identifying the stocks that would enter the baseline long-only momentum portfolio for the following month. Specifically, we consider those in the top decile of past 12-month returns, excluding the most recent month. To broaden the coverage of the LLM assessment, we extend this set to also include stocks from the second decile, resulting in a wider pool of momentum candidates.
    
    For each stock in this extended set, we collect news data shortly before the market close. We gather all news articles published from 15:45 of the previous $k$ business days (\(t-k\)) to 15:45 of the current day (\(t\)), where the look-back period $k$ is a parameter to be optimized.
    This set of news for each stock-day pair is then processed by a large language model (LLM) using a custom-designed prompt. The prompt performs three key functions: it describes the logic of the baseline momentum strategy, informs the model that the stock has been selected based on momentum criteria, and requests a score that reflects the likelihood of the stock continuing to perform well in the near future. The LLM outputs a score between 0 and 1, representing its assessment based on the content of the news. Scores are then normalized to range between \(-1\) and \(+1\).

    The stocks in the extended set are subsequently ranked according to their LLM scores. In cases where multiple stocks receive identical scores, the original momentum ranking is used to break the ties. From this final ranked list, the top \(m\) stocks are selected to form the enhanced portfolio. The parameter \(m\) is treated as a hyperparameter and is subject to optimization.

    Initially, all selected stocks are assigned baseline weights, whereas the baseline weighting strategy is either equal- or value-weighted. To incorporate the strength of the LLM signal into the allocation, we then tilt these weights based on each stock's score. The final weight \(W_i\) of stock \(i\) is computed as follows:
    \[
    W_i = E \cdot \eta^{\text{LLM\_Score}_i}
    \]
    where \(E\) is the baseline weight, \(\text{LLM\_Score}_i\) is the normalized score of stock \(i\), and \(\eta\) is a scaling parameter to be optimized. 
    After $W_i$ are computed for all stocks $i$, the weights are standardized to sum-up to one. 
    This procedure enables the LLM to influence both stock selection and capital allocation, adding a layer of information-driven refinement to the traditional momentum strategy.

\subsection{Sample Splitting and Hyperparameter Selection}

    To evaluate the performance of the LLM-enhanced momentum strategy in a robust and realistic manner, we split the sample, which encompasses with 1,382 daily observations, into two distinct subsamples. 
    A \textbf{Validation Set} covering the period from October 2019 to December 2023, 
    and a \textbf{Test Set} spanning from January 2024 to March 2025.
    Table~\ref{tab:valtest} summarizes the two periods and reports the number of daily observations for each subset.

    The Validation Set is used to identify the best-performing hyperparameter configuration for the LLM-enhanced strategy. Specifically, we search over a grid of candidate values for parameters such as the LLM lookback window $k$, the look-ahead window $l$, the number of selected stocks $m$, and the score-based weighting multiplier $\eta$. 
    Further, parameters include 
    the prompt category $\pi$ described in Section \ref{sub:prompts}, 
    the rebalancing frequency $\tau$ (monthly or weekly),
    a boolean variable $c$ specifying whether a weight constraint of maximum 15\% per stock is applied, 
    a dummy $w$ indicating if the initial weighting scheme is equally-weighted or value-weighted, and 
    the number of stocks $m$ included in the portfolio, selected through the LLM scoring detailed above.
    For each parameter combination 
    $\theta = ( 
        \tau,   
        k,      
        l,      
        m,      
        \pi,    
        c,      
        w,      
        \eta    
    )$, 
    we compute portfolio returns and evaluate the following objective function:
    \begin{equation}
        U( \theta ) = \frac{3}{4} \, \text{Sharpe}(\theta) - \frac{1}{4} \, \text{MDD}(\theta)
    \end{equation}
    Here, $\text{Sharpe}(\theta)$ and $\text{MDD}(\theta)$ represent the percentile ranks (between 0 and 1) of the Sharpe ratio and maximum drawdown across all evaluated parameter sets. This formulation places greater weight on risk-adjusted performance while penalizing downside risk. Our findings are qualitatively robust to alternative weightings in the objective function.
    Once the optimal parameter set $\theta^*$ is identified based on the validation data, we evaluate its performance out-of-sample on the Test Set. This procedure provides a realistic estimate of future strategy performance, simulating a situation where the model and parameters are fixed in advance. The Test Set results are thus unaffected by overfitting to past data.

    Importantly, our setup does not require a dedicated “training set” because the language model used, ChatGPT 4.0 mini, is already pre-trained. We do not perform any fine-tuning of the model, primarily due to the relatively limited size of our dataset. Nevertheless, incorporating task-specific fine-tuning represents an interesting direction for future research. In our implementation, the model is treated as a static, black-box interpreter of textual input, and its parameters remain fixed throughout the study.

    Moreover, since ChatGPT 4.0 mini was trained on data only up to October 2023, the Test Set, which extends from January 2024 to March 2025, lies in a truly out-of-sample regime even with respect to the model's own knowledge base. In particular, results from November 2023 onward are guaranteed to be free from any form of information leakage via model pre-training. 
    While \cite{he2025chronologically} suggest that information leakage is limited in financial prediction tasks, our approach makes the evaluation particularly rigorous, and it isolates the model's ability to generalize from its training data rather than relying on memorized historical information.

    \begin{table}[t]
      \centering
      \begin{tabular}{lrrrr}
      \textbf{Subsample} & \textbf{Start} & \textbf{End} & \textbf{Length} & \textbf{Observations} 
      \\[0.3em]
      \midrule
      \\[0.0em]
      \textbf{Validation Set} & October 2019  & December 2023 & $\sim$ 4 years & 1070 \\[1.3em]
      \textbf{Test Set}       & January 2024  & March 2025    & 15 months      & 312  \\[0.8em]
      \bottomrule
      \end{tabular}
      \caption{Description of Validation and Test Sets}
        \label{tab:valtest}
    \end{table}

\subsection{Prompts}\label{sub:prompts}

    We apply prompt engineering techniques to interact with state-of-the-art ChatGPT models developed by OpenAI. The goal of each prompt is to extract an informative signal from a sequence of firm-specific news items, indicating whether the stock is likely to continue its recent price trend. To do so, the prompt explicitly informs the model that the stock in question is about to enter a momentum portfolio, thereby framing the analysis as a forward-looking investment decision. By doing so, the model is encouraged to evaluate the content and tone of the news in the context of momentum continuation, identifying whether recent information supports or contradicts the trend implied by past returns.

    We construct two distinct categories of prompts that vary in complexity and structure: a \emph{Basic} prompt and a more \emph{Advanced} one.
    The \emph{Basic Prompt} includes a brief mention of the baseline momentum strategy and provides a concise task description focused on the final objective, that is, predicting the likelihood of return continuation. It includes minimal context about the portfolio formation process and does not impose any explicit structure on the news feed. The goal is to assess whether a more lightweight formulation can still generate useful signals from the model.
    The \emph{Advanced Prompt}, instead, provides a more detailed and structured formulation. It includes an explicit explanation of how the baseline momentum portfolio is formed, describes the forecast horizon, and outlines how the model should interpret the news items. In particular, it offers suggestions on the types of information that may be relevant for predicting price continuation and describes the organization of the news feed provided within the prompt.

    All prompts are dynamically generated and tailored to the specific stock and evaluation context. In particular, each prompt depends on: (i) the stock ticker under consideration, (ii) the look-back window $k$ used to gather news, and (iii) the forecast horizon for the momentum signal. 
    At the end of each prompt, a list of stock-specific news articles released over the previous $k$ business days is appended. For each article, we include the title, a short summary, and the time elapsed since publication, expressed relative to the request timestamp. This format enables the model to account for both the content and recency of the news.
    The prediction horizon is a function of the rebalancing frequency: 5 days for weekly rebalancing and 21 days for monthly rebalancing.
    Examples of both the basic and advanced prompts are provided below. These are dynamically generated based on the stock ticker (AAPL), the look-back window (5 days), and the forecast horizon (21 days). 
    The examples below do not include the list of the actual stock-specific news items, which are automatically appended at the end of the prompt.

    \subsection*{Basic Prompt Example}
    \begin{lstlisting}
    ---- Introduction----
    You are a financial analysis model tasked with constructing and rebalancing a long-only 
    S&P 500 portfolio based on momentum signals and news-driven sentiment scores for each stock.
    
    ---- Context Preamble ----
    At the end of each month, you form a momentum portfolio of the stocks of the S&P 500. 
    To enhance the baseline momentum portfolio, you will also analyze news sentiment. 
    Each EM, the portfolio is rebalanced based on the news of each title. 
    
    ---- News Block Information----
    Below is a list of news related to the US Equity stock with ticker AAPL, released over 
    the previous 5 business days. They are ordered by release time, which is reported in 
    square brackets. 
    
    ---- Task Description ----
    Your task is to read all news headlines and summaries for AAPL published up to 
    2024-01-03 15:55 (NYSE time), and predict whether the stock's upward momentum will 
    continue through the close on 2024-02-01 (21 business day(s) later).
    
    ---- Output Format ----
    The prediction is expressed as a score from 0 to 1, where 0 means the momentum will stop 
    or revert, and 1 indicates that momentum will continue. Intermediate values (e.g., 0.5615) 
    represent the probability you assign to momentum continuation.
    
    
    The format of the news is the following:
    
    [5 hours ago]
    Title
    Short Summary
    
    [8 hours ago]
    Title
    Short Summary
    
    **IMPORTANT**: Output only the score (e.g., `0.7341`) and nothing else.
    \end{lstlisting}
    
    \subsection*{Advanced Prompt Example}
    \begin{lstlisting}
    ---- Introduction ----
    You are a financial analysis model tasked with constructing and rebalancing a long-only 
    S&P 500 portfolio based on momentum signals and news-driven sentiment scores for each stock.
    
    ---- Context Preamble ----
    Your baseline strategy is a long-only portfolio of US equity stocks based on the Carhart 
    momentum factor. At the end of each month, the portfolio includes the top decile of stocks 
    ranked by past-year returns (excluding the last month), under the assumption that recent 
    good performing stocks continue their trend. To enhance the baseline momentum portfolio, 
    you will also analyze news sentiment. Specifically, at the last business day every EM, you 
    rebalance the portfolio by computing each stock's news-based sentiment score to assess 
    their future performance. Currently, you're working with the stock AAPL. 
    
    ---- News Block Information ----
    To assess if we can improve the baseline portfolio, you are provided with a series of news 
    headlines and summaries of the main information related to AAPL, published between 
    **2024-01-03 15:55 (NYSE time)** - just before the market close on the reference date 
    2024-01-03. These news items are time-stamped and sorted from most recent to oldest. 
    Each item has the timestamp in brackets, then a Title line, then ">>>", then a Short Summary line. 
    
    ---- Task Description ----
    Your task is to assess whether the current upward momentum in price of the stock AAPL is 
    supported by the existing news headlines and summaries, and whether the stock is likely 
    to continue rising until the next date - **2024-02-01**, the close of trading 21 business 
    day(s) from the reference date. To do this, you read all news articles of the stock, 
    evaluate their sentiment and assess the likelihood that the stock will outperform the other 
    stocks in the S\&P 500 universe. Follow these instructions: Carefully interpret the news 
    content for its short-term market implications. Assign greater weight to more recent news, 
    especially within the last 24-48 hours. Judge whether the overall tone and relevance of 
    the news suggest continued upward momentum. At the end, provide your confidence score 
    that AAPL will continue its upward trend until 21 business day(s) from the reference date.
    
    ---- Output Format ----
    The prediction is expressed as a score from 0 to 1, where 0 means the momentum will stop 
    or revert, and 1 indicates that momentum will continue. Intermediate values (e.g., 0.5615) 
    represent the probability you assign to momentum continuation.
    
    
    The format of the news is the following:
    
    [5 hours ago]
    Title
    Short Summary
    
    [8 hours ago]
    Title
    Short Summary
    **IMPORTANT**: Output only the score (e.g., `0.7341`) and nothing else.
    \end{lstlisting}

\section{Results}

    This section presents the main empirical findings of the paper. We begin by reporting summary statistics for the strategy components and the LLM-generated signals. This provides a descriptive view of the underlying data used to construct and evaluate the LLM-enhanced momentum portfolio.
    Next, we compare the performance of the LLM-enhanced strategy to the baseline momentum portfolio. We document the set of optimized hyperparameters selected using the validation set and examine the behavior of the strategy over time, focusing on both in-sample and out-of-sample periods. Performance metrics are evaluated at the daily frequency to provide a high-frequency view of return dynamics.
    Finally, we conduct a detailed parameter analysis to investigate how the performance of the strategy varies with different choices of key hyperparameters. This helps to understand the robustness of our results and sheds light on the economic role played by each modeling choice.

\subsection{Summary Statistics}

    \begin{figure}[t]
        \centering
        \begin{subfigure}[t]{0.5\textwidth}
            \centering
            \includegraphics[width=1\textwidth]{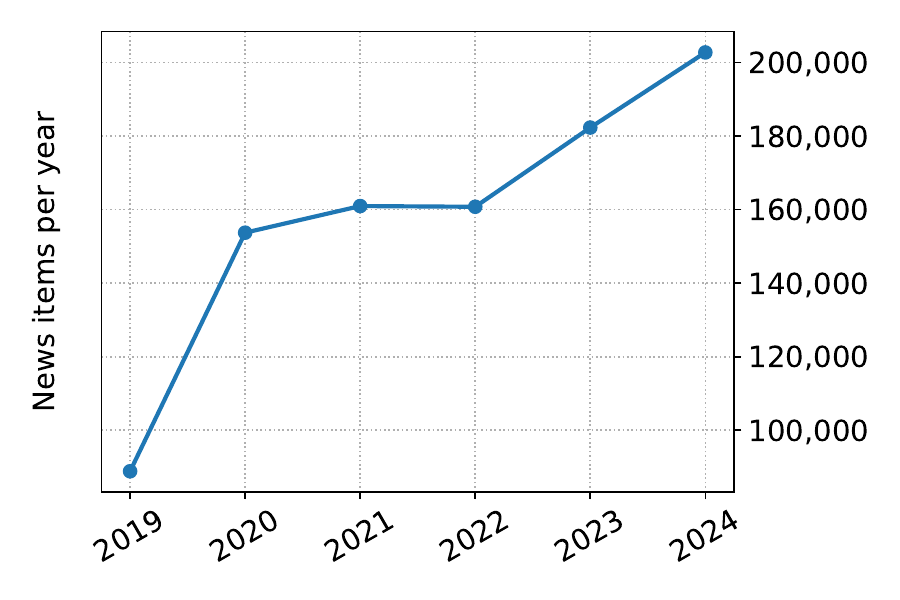}
            \caption{Time-Series of total yearly news items}
        \end{subfigure}%
        ~ 
        \begin{subfigure}[t]{0.5\textwidth}
            \centering
            \includegraphics[width=\textwidth]{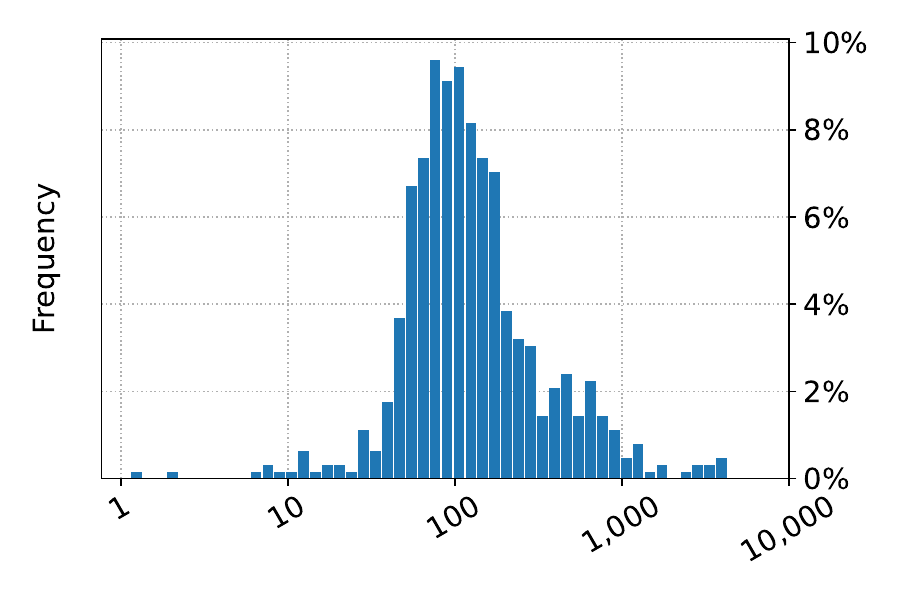}
            \caption{Number of stock-specific news per year}
        \end{subfigure}\vspace{1em}

        \begin{subfigure}[t]{0.5\textwidth}
            \centering
            \includegraphics[width=\textwidth]{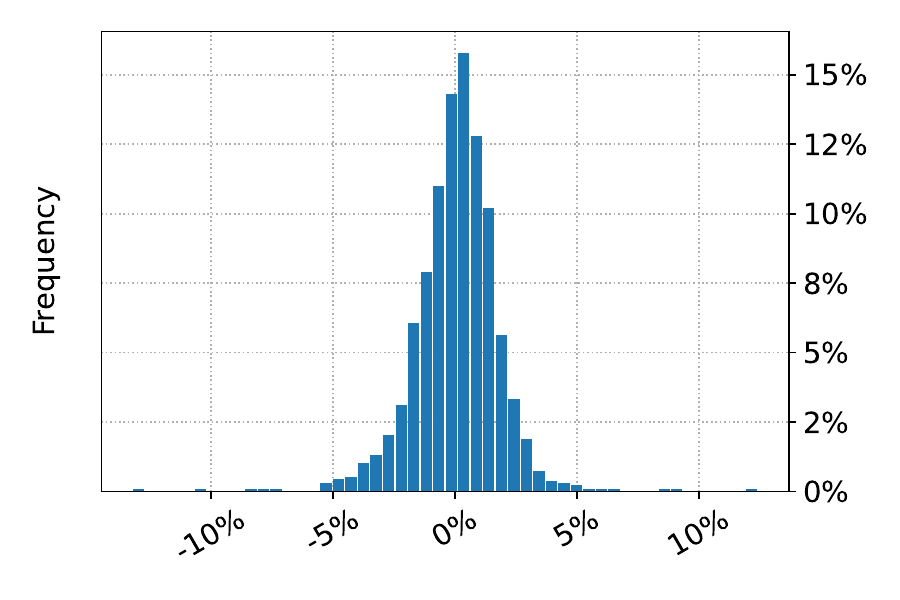}
            \caption{Distribution of daily returns generated by the baseline momentum portfolio}
        \end{subfigure}%
        ~ 
        \begin{subfigure}[t]{0.5\textwidth}
            \centering
            \includegraphics[width=\textwidth]{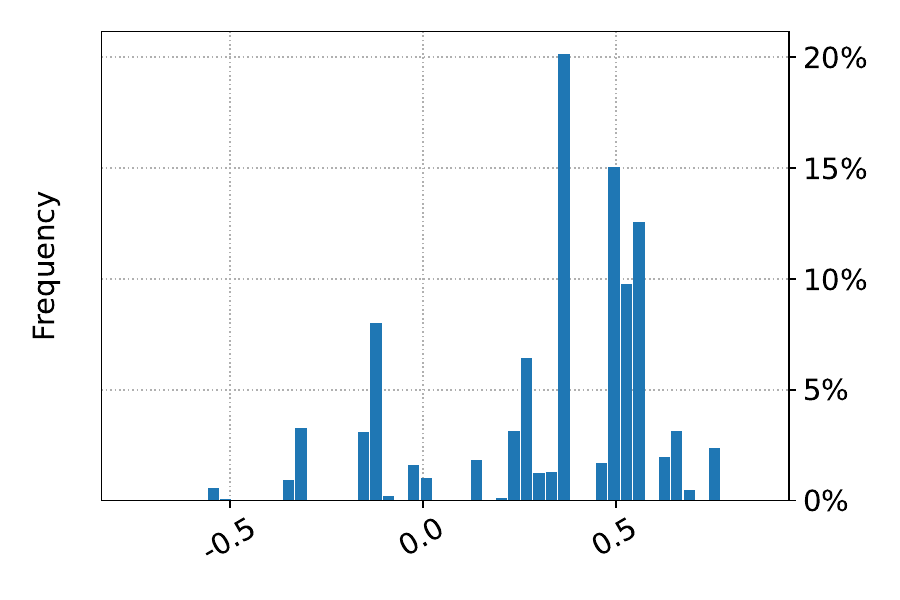}
            \caption{Distribution of LLM scores at the stock-day level}
        \end{subfigure}
        \caption{Summary Statistics}
        \label{fig:stats}
    \end{figure}

    Figure~\ref{fig:stats} provides a set of summary statistics describing the key components of our dataset and strategy inputs.
    Panel (a) displays the total number of news articles published each year from 2019 to 2024. The data reveal a marked upward trend, with yearly news volume increasing from under 100{,}000 items in 2019 to over 200{,}000 in 2024. This growth reflects both improved coverage of stock-specific news and increasing activity from automated financial media outlets. The rising trend reinforces the relevance of scalable automated methods for real-time news processing in asset management.

    Panel (b) presents the distribution of the number of stock-specific news articles per firm-year. The x-axis is displayed on a logarithmic scale. As expected, the distribution is heavily right-skewed: while the median firm receives a few hundred articles per year, a small number of high-profile companies attract several thousand. This skewness highlights the disproportionate media attention given to mega-cap firms and underscores the importance of a model capable of processing unbalanced textual input across the cross-section.

    Panel (c) shows the distribution of daily returns for the baseline long-only momentum portfolio. The distribution is approximately symmetric and centered near zero, with fat tails capturing the presence of both large positive and negative return days. This shape is consistent with known properties of daily equity returns, even after applying systematic selection filters such as momentum.

    Panel (d) reports the distribution of LLM-generated scores at the stock-day level. These scores are obtained by applying either the basic prompt to the news feed for each stock, and they are normalized to lie between \(-1\) and \(+1\). To simplify the visualization, the plot excludes scores that are missing due to the absence of relevant news on a given day. In the actual empirical analysis, those missing scores are treated as zero, reflecting a neutral stance in the absence of new information. The distribution exhibits significant variation across the cross-section and over time, which is critical for generating differentiated trading signals.

    Taken together, these summary statistics confirm the richness and heterogeneity of the textual dataset, as well as the plausibility of the LLM-generated scores as a conditioning signal for systematic investment strategies.

\subsection{Optimal Hyperparameters}\label{sub:optimal_hyperparameters}

    We begin by identifying the optimal configuration of hyperparameters using the validation set and then assess the performance of the resulting strategy both in-sample and out-of-sample. The goal is to determine whether incorporating LLM-generated signals can yield economically and statistically significant improvements in return characteristics while preserving robustness and stability.

    A total of 512 combinations of hyperparameters $\{\theta_i\}_{i=1,\dots,512}$ is tested on the validation set. For each combination $\theta_i$, the objective function $U(\theta_i)$ is evaluated, and the best model is selected as the top performing one according to that metric. Table~\ref{tab:optimal-params} provides a summary of the possible values for each parameter and the resulting optimal parameter set $\theta^*$.

    The optimal strategy rebalances monthly ($\tau = \text{month}$), which reflects a trade-off between responsiveness to new information and transaction cost efficiency. 
    
    The lookback window for news retrieval is set to one day ($k = 1$), indicating that the most recent firm-specific news carries the strongest predictive power for short-term return continuation. The look-ahead window is set to 21 days ($l = 21$), although there is no notable difference in the prediction quality regarding their look-ahead horizons.

    The optimal number of stocks selected each month is $m = 50$, suggesting that the LLM-enhanced signal is most effective when applied to a relatively concentrated set of high-conviction positions. 
    
    The best-performing prompt is the simpler of the two alternatives ($\pi = \text{basic}$), suggesting that a concise task description is sufficient to extract useful signals from the model in this setting.

    The initial allocation is market-cap weighted across selected stocks ($w = \text{Value-weighted}$), suggesting that the performance of the LLM-enhanced portfolio can best possibly be leveraged when the model builds on the stronger informational foundation of large companies. This is somewhat intuitive, as larger firms typically generate more news coverage and market commentary, providing the LLM with richer data inputs from which to learn and extract signals.
    Further, the LLM scores are used to tilt these weights using a multiplier of $\eta = 5$. This relatively high multiplier indicates that the cross-sectional variation in LLM scores is meaningful and can be confidently exploited for capital allocation.

    \begin{table}[t]
        \centering
        \begin{tabular}{lllr}
        \toprule
        \textbf{Symbol}$\qquad$ & \textbf{Parameter} & \textbf{Options} & \textbf{Optimal Value} \\
        \midrule
        $\tau$   & Rebalancing frequency     & Day, Month & Month \\
        $k$      & Lookback window           & 1 day, 5 days & 1 day \\
        $l$      & Lookahead window          & 1 day, 21 days & 21 days \\
        $\pi$    & Prompt type               & Basic, Advanced & Basic \\
        $m$      & Number of stocks          & 25, 50, 75, 100 & 50 \\
        $c$      & Weights constraints       & True, False & True \\
        $w$      & Initial weighting scheme  & Equal, Value & Value \\
        $\eta$   & Weight multiplier         & 1.25, 2.5, 3.75, 5 & 5 \\
        \bottomrule
        \end{tabular}
        \caption{Optimal hyperparameter configuration selected on the validation set}
        \label{tab:optimal-params}
    \end{table}

    Overall, the optimal configuration reflects a disciplined implementation, a focus on the most recent news, and an emphasis on strong conviction signals within a diversified yet selective portfolio.

\subsection{Performance of LLM-Enhanced Strategy}

    The LLM-enhanced momentum strategy is evaluated against its natural benchmark, the baseline momentum portfolio. Both strategies are described in Section~\ref{sec:methodology}. To ensure comparability, the baseline universe and the rebalancing frequency are the same for both portfolios, leaving the performance differences solely to the selection (ranking) and allocation (tilt) effects of the LLM model. The LLM model, together with its hyperparameters, is selected based solely on Validation Set performance. This in-sample selection procedure helps mitigate the risk of overfitting, ensuring that the reported out-of-sample results provide a realistic assessment of the model's predictive ability.

    Table~\ref{tab:perf-stats} reports the main performance statistics for both the full sample and the out-of-sample period. In the full-sample evaluation, the LLM-enhanced portfolio exhibits a Sharpe ratio of 0.69, compared to 0.57 for the baseline. The Sortino ratio, which emphasizes downside risk, improves from 0.54 to 0.69. The annualized return increases from 15\% for the baseline to 18\% for the LLM-enhanced version. Volatility decreases slightly from 26\% to 24\%, while the maximum drawdown improves from $-33\%$ to $-31\%$. 
    
    As is to be expected, the performance improvement comes at the expense of a larger turnover, indicating that the LLM-enhanced strategy incurs higher trading activity. However, because all reported results are net of transaction costs, the increased turnover is offset by the stronger predictive signals, meaning that the strategy’s added value persists even after accounting for trading frictions.

    \begin{table}[t]
        \centering
        \begin{tabular}{lcccc}
        \toprule
        & \multicolumn{2}{c}{\textbf{Full Sample}} & \multicolumn{2}{c}{\textbf{Out-of-Sample}} \\
        \cmidrule(lr){2-3} \cmidrule(lr){4-5}
        Metric & Baseline & LLM-Enhanced & Baseline & LLM-Enhanced \\
        \midrule
        Sharpe          & 0.57  & 0.69  & 0.79  & 1.06 \\
        Sortino         & 0.54  & 0.69  & 0.93  & 1.28 \\
        Return          & 0.15  & 0.18  & 0.24 & 0.30 \\
        Volatility      & 0.26  & 0.24  & 0.24  & 0.22 \\
        MDD             & -0.33  & -0.31  & -0.19  & -0.17 \\
        Turnover        & 0.62  & 0.90  & 0.48  & 0.80 \\
        \bottomrule
        \end{tabular}
        \caption{The Table presents the performance statistics of the baseline and LLM-enhanced momentum strategies for the full sample and out-of-sample period, assuming transaction costs of 2 bps. The Sharpe and Sortino ratios, the return and the volatility are annualized. }
        \label{tab:perf-stats}
    \end{table}

    In the out-of-sample period, which starts in January 2024, the improvements are even more pronounced. The Sharpe ratio improves from 0.79 for the baseline to 1.06 for the LLM-enhanced strategy, while the Sortino ratio increases from 0.93 to 1.28. The total return increases from approximately 24\% to 30\%, with lower volatility (22\% versus 24\%) and a smaller maximum drawdown ($-17\%$ compared to $-19\%$). Turnover results remain similar, fostering the claim that the incurred excess costs due to increased trading frequency are offset by stronger predictive signals. In other words, the LLM-enhanced approach delivers a substantial improvement in risk-adjusted returns over the baseline strategy.

    While these results are encouraging, they should be interpreted with caution given the limited size of the out-of-sample period. For example, a time-series regression of LLM-enhanced returns on baseline returns yields a positive annualized alpha of 3.26\%, but the estimate is statistically significant only at the 10\% level. Nevertheless, it is worth noting that all results are computed under the assumption of 2 basis points of transaction costs per trade, which is a conservative estimate given the high liquidity of S\&P 500 stocks in our sample.

\begin{figure}[t]
        \centering
        \includegraphics[width=\textwidth]{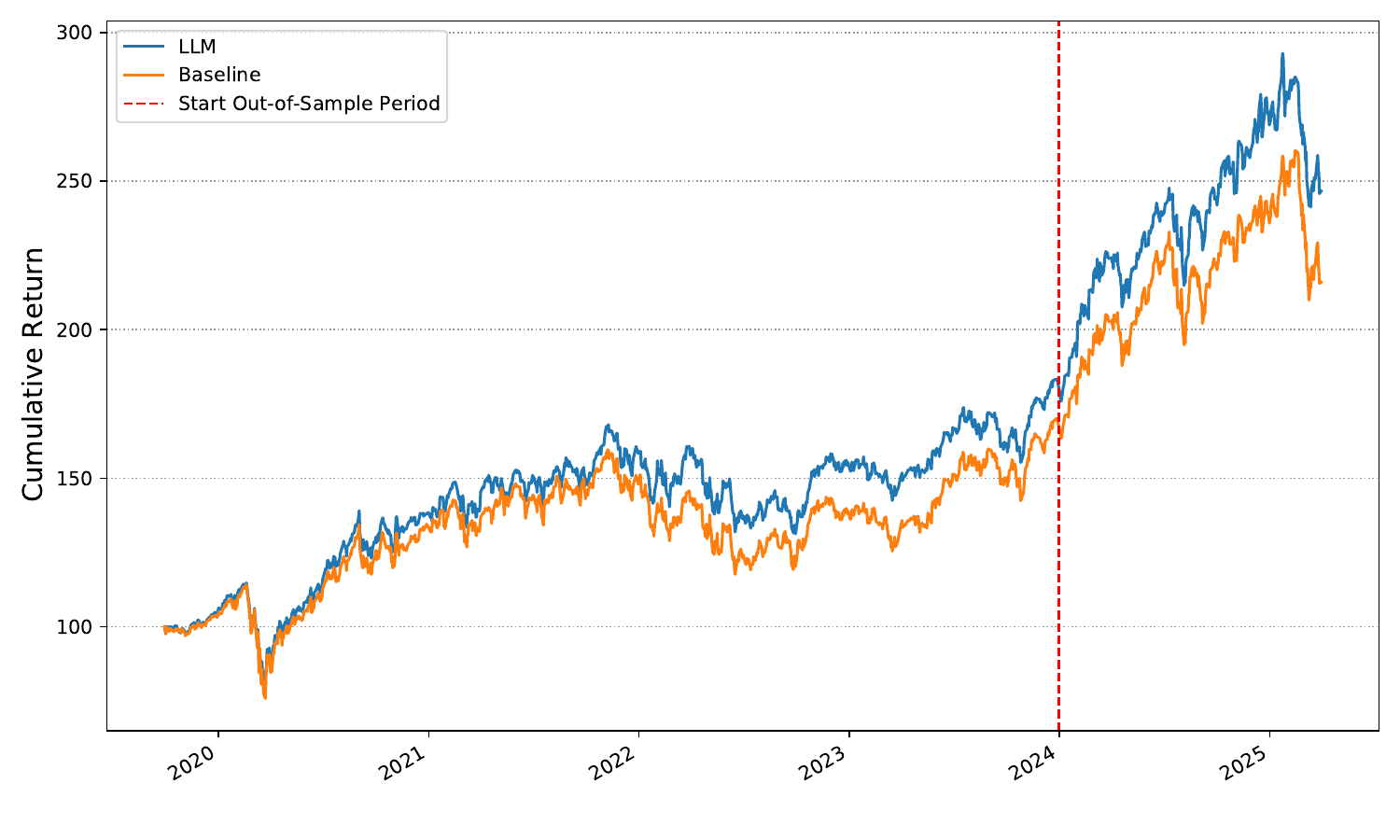}
        \caption{The Figure displays cumulative returns at a daily frequency of the best performing LLM-Enhanced strategy, based on in-sample parameters optimization (blue line), and the baseline momentum strategy (orange line). 
        The vertical dashed line indicates the start of the out-of-sample period.}
        \label{fig:cumrets}
    \end{figure}

    Figure~\ref{fig:cumrets} plots the cumulative returns of the two strategies over time. The LLM-enhanced portfolio outperforms the baseline consistently throughout the sample, with notable divergence in performance after 2023. The vertical dashed line marks the start of the out-of-sample period in January 2024. This period is particularly important because it begins after the October 2023 pre-training cut-off of ChatGPT 4.0 mini. Consequently, the LLM’s performance in this phase cannot be attributed to memorization or information leakage from its training corpus. The fact that the LLM-enhanced strategy continues to outperform substantially during this period supports the conclusion that the gains stem from the model's real-time ability to interpret news and enhance momentum signals, rather than from prior knowledge embedded in the pre-training stage.

    Overall, the evidence indicates that incorporating LLM-generated signals into a standard momentum framework can lead to economically meaningful improvements in performance, both in-sample and out-of-sample. The gains are achieved with lower levels of volatility, and under conservative assumptions for transaction costs of 2 bps. While the limited out-of-sample period calls for cautious interpretation, the consistency of the results across sample splits and the stronger performance after the LLM’s pre-training cut-off suggest that the approach captures genuine incremental information from firm-specific news. These findings provide a compelling case for the integration of advanced language models into systematic equity strategies, and motivate further research into refining prompt design, model selection, and integration with other factor signals.


\subsection{Parameters Perturbation Analysis}

    \begin{figure}
        \centering
        \begin{subfigure}[t]{0.49\textwidth}
            \centering
            \includegraphics[width=\textwidth]{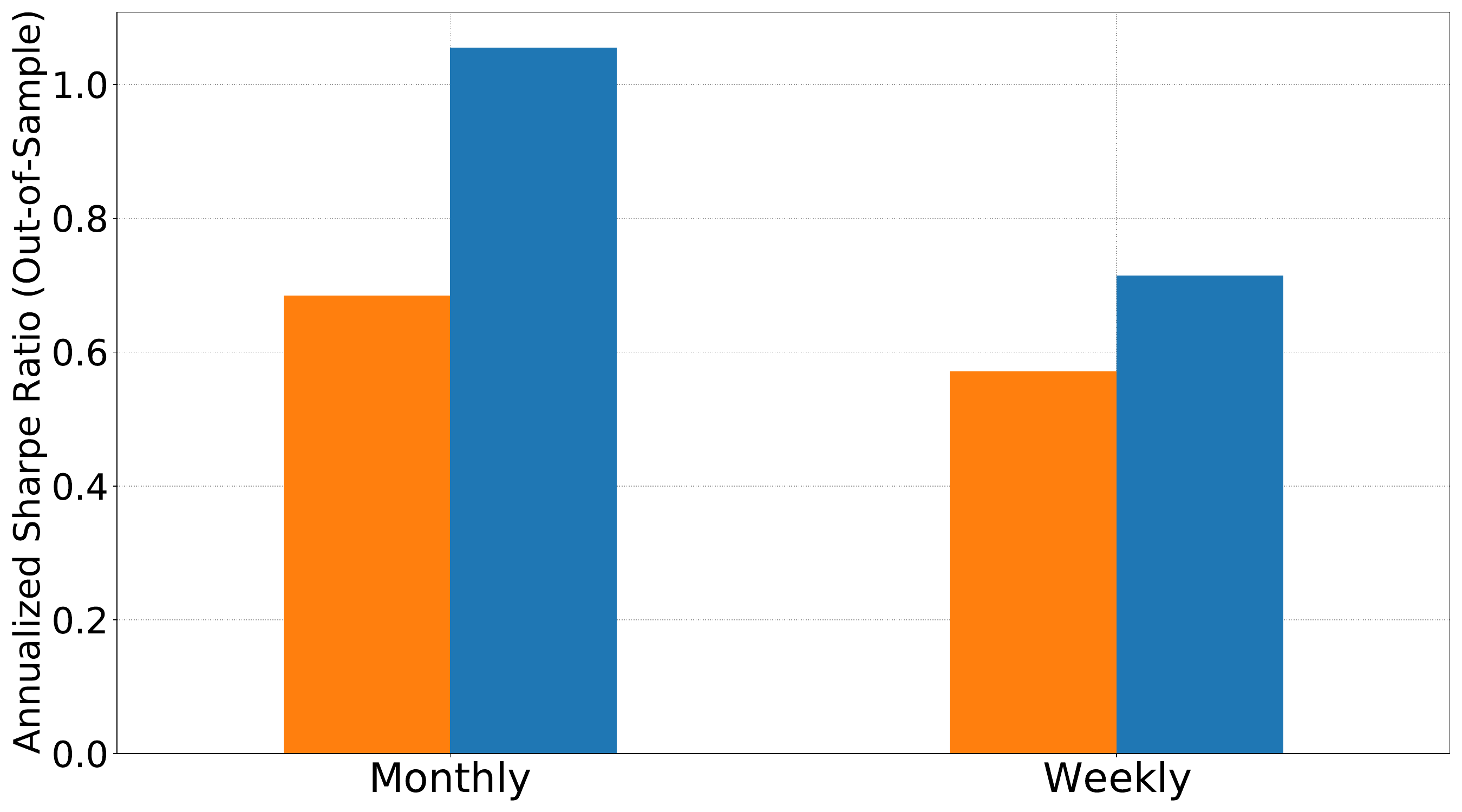}
            \caption{Sharpe ratio by rebalancing frequency}
        \end{subfigure}%
        ~ 
        \begin{subfigure}[t]{0.49\textwidth}
            \centering
            \includegraphics[width=\textwidth]{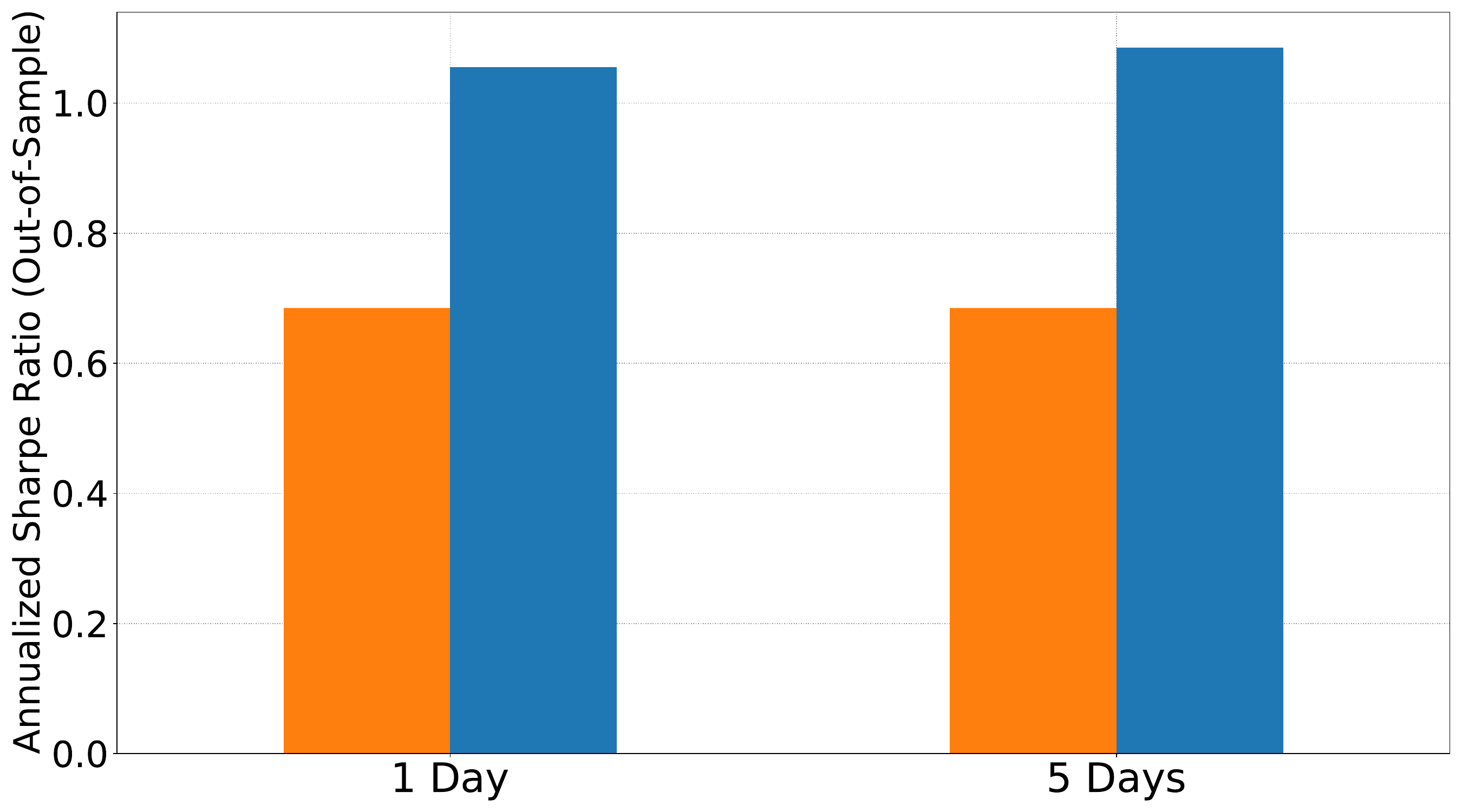}
            \caption{Sharpe ratio by backward-looking window}
        \end{subfigure}
        \par\bigskip\par\bigskip
        \begin{subfigure}[t]{0.49\textwidth}
            \centering
            \includegraphics[width=\textwidth]{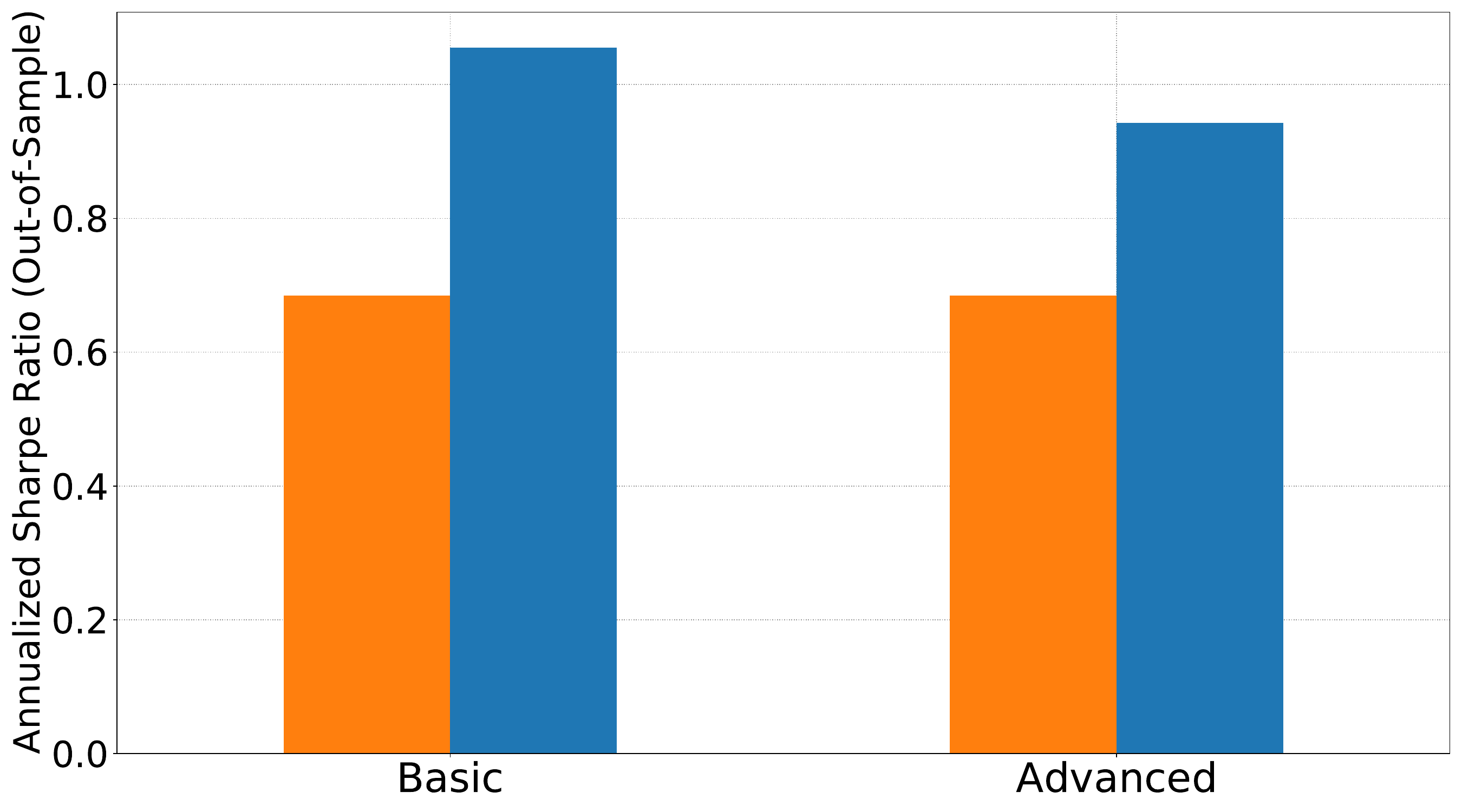}
            \caption{Sharpe ratio by LLM prompt specification}
        \end{subfigure}%
        ~ 
        \begin{subfigure}[t]{0.49\textwidth}
            \centering
            \includegraphics[width=\textwidth]{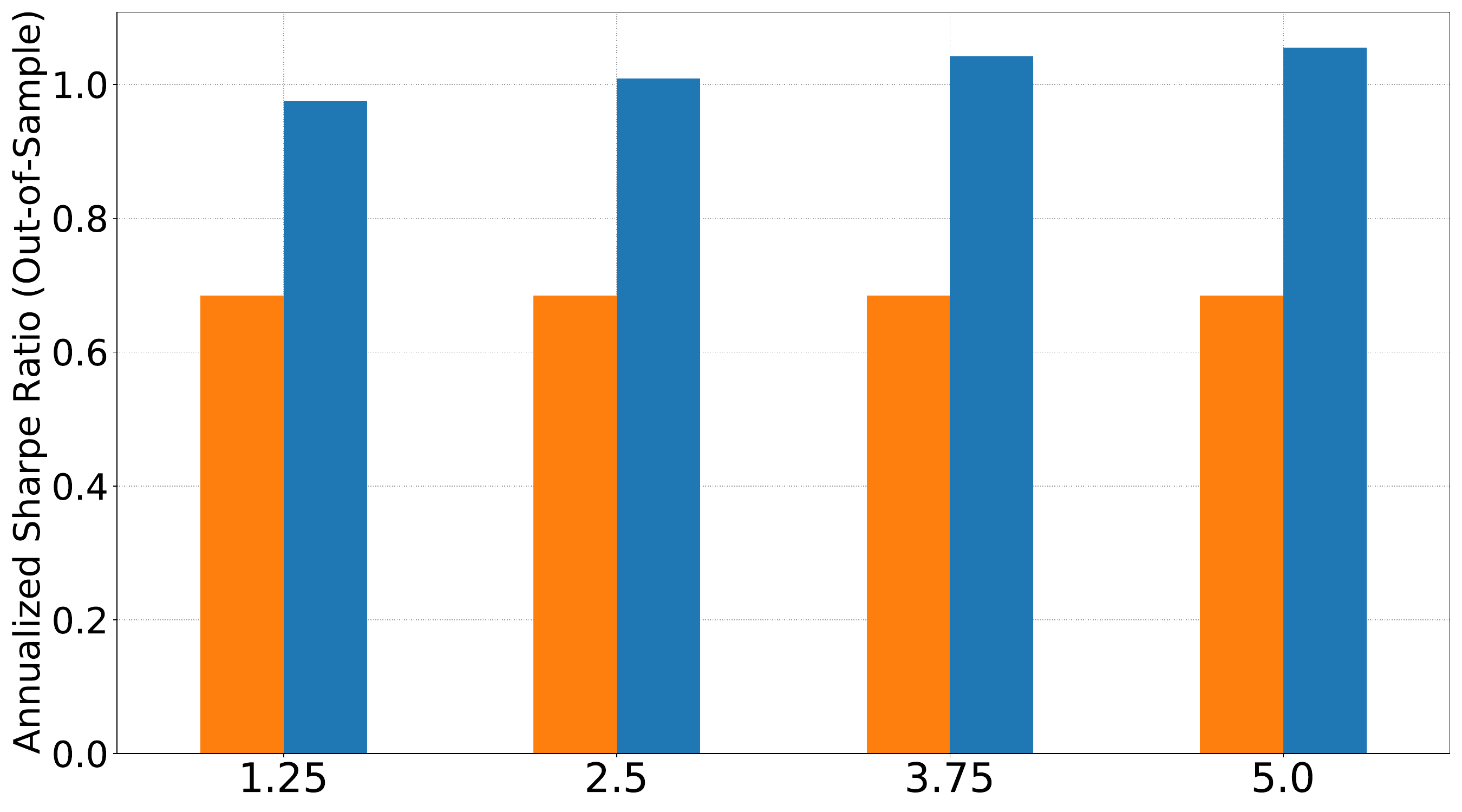}
            \caption{Sharpe ratio by tilt factor}
        \end{subfigure}
        \par\bigskip\par\bigskip
        \begin{subfigure}[t]{0.49\textwidth}
            \centering
            \includegraphics[width=\textwidth]{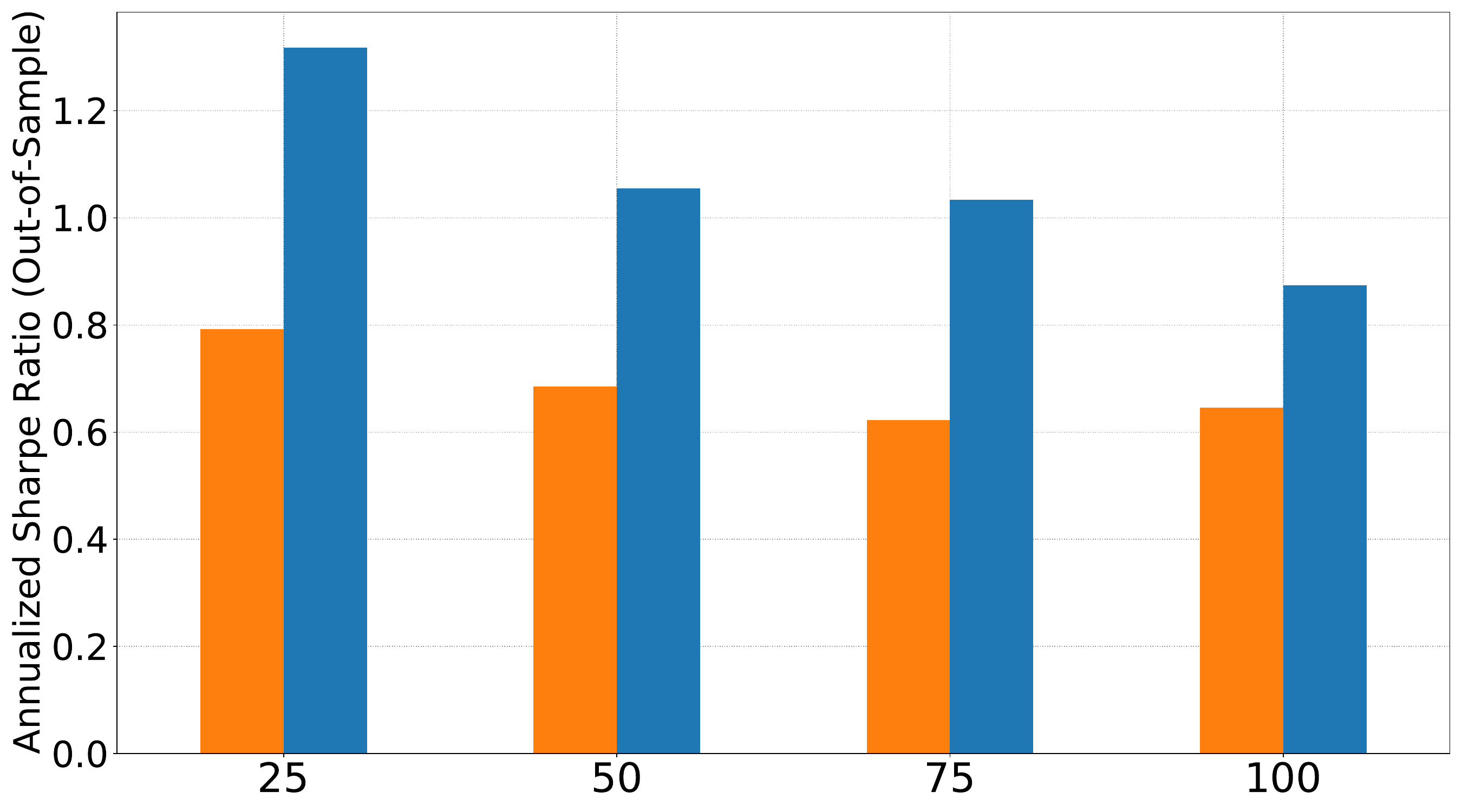}
            \caption{Sharpe ratio by number of stocks}
        \end{subfigure}%
        ~ 
        \begin{subfigure}[t]{0.49\textwidth}
            \centering
            \includegraphics[width=\textwidth]{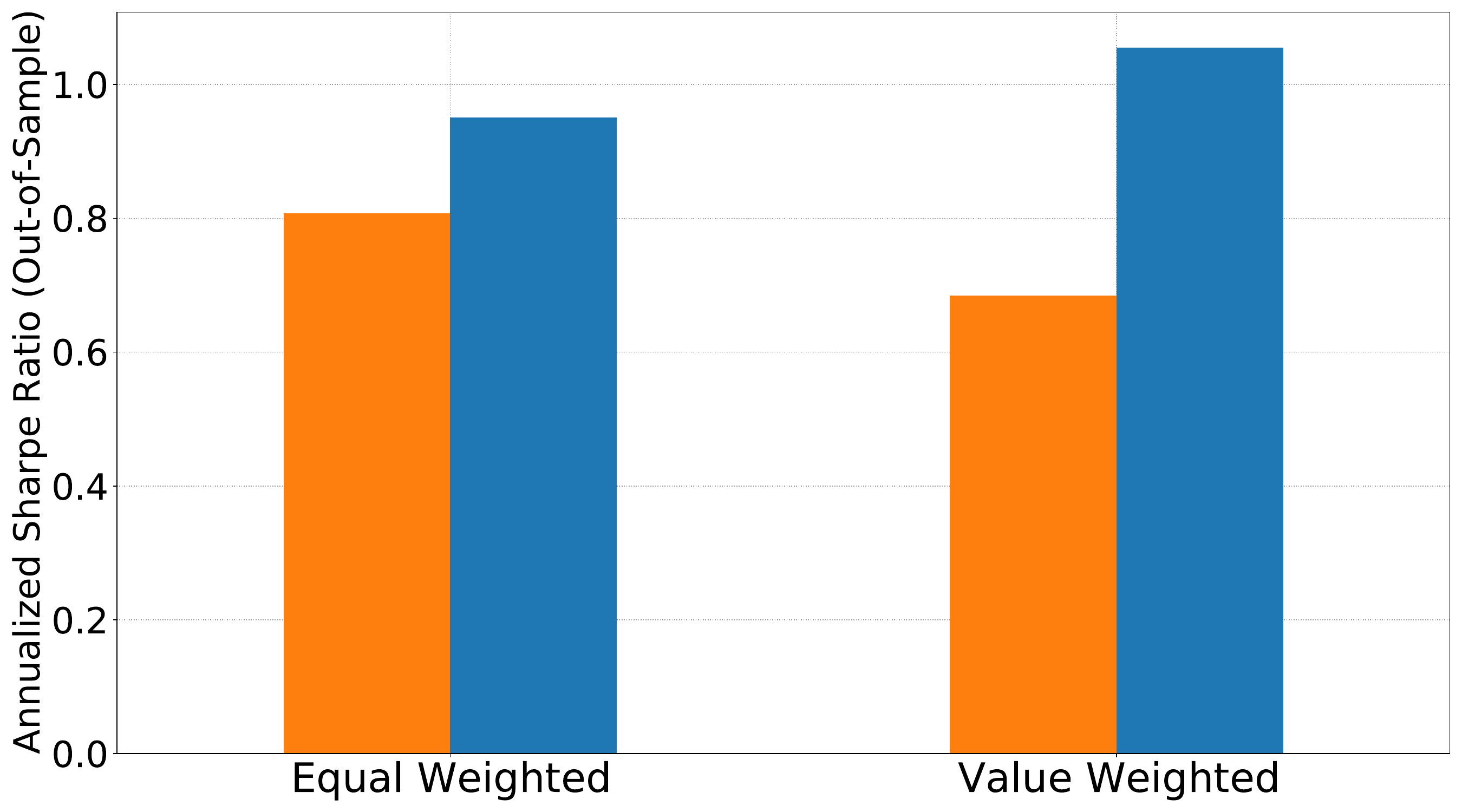}
            \caption{Sharpe ratio by baseline weighting strategy}
        \end{subfigure}
        \vspace{1em}
        \caption{Ceteris Paribus Perturbation Analysis. The blue bars indicate the sharpe ratio of the LLM strategy. The orange bars the sharpe ratio of the baseline strategy.}
        \label{fig:para}
    \end{figure}
    
    In this section we investigate how the performance of the LLM-enhanced strategy relates to changes in key parameters. To this end, we adopt a \textit{ceteris paribus perturbation analysis} inspired by sensitivity analysis techniques in the machine learning literature \citep{feurer2019automated,biecek2021explanatory}. Specifically, for each parameter $p$ of interest, we start from the optimal configuration $\theta^*$ identified in Section~\ref{sub:optimal_hyperparameters} and vary $p$ across all of its admissible values while keeping all other parameters fixed at their optimal settings. This approach can be viewed as a local, one-dimensional projection of the performance surface around $\theta^*$, which isolates the marginal effect of each parameter on strategy performance.
    We focus on out-of-sample results for this exercise, as starting from $\theta^*$, which is itself chosen based on the validation set, provides a natural basis for evaluating the stability of the model in a realistic forward-looking setting. Nevertheless, the qualitative patterns we discuss here are also observed when using the full sample.

    Figure~\ref{fig:para} reports the annualized out-of-sample Sharpe ratio of the LLM-enhanced strategy and the baseline momentum portfolio for different parameter values. Several findings emerge:
    Panel (a) shows that the LLM-enhanced strategy performs remarkably better at a monthly rebalancing frequency than at a weekly one, with Sharpe ratios of approximately 1.1 and 0.7, respectively. This suggests that the cost of higher turnover outweighs the benefits of more frequent signal updates. The baseline strategy displays a similar pattern, although with lower Sharpe ratios across both frequencies. 

    Panel (b) investigates the impact of the amount of news fed to the model. Expanding the input window from 1 to 5 days of news only marginally improves the Sharpe ratio. This suggests that more news does not necessarily translate into better signals, as markets tend to incorporate new information into prices rather quickly. This is a useful insight, as it indicates that similar results can be achieved with less data, which in turn helps to reduce the costs associated with deploying LLMs in practice.

    Panel (c) compares the two prompt types, revealing that the basic prompt outperforms the advanced version, with out-of-sample Sharpe ratios of roughly 1.1 and 0.95, respectively. This indicates that the simpler formulation may be sufficient to guide the LLM’s interpretation of the news feed in this context. The difference is not large enough to be statistically significant, suggesting that both designs are viable. 

    Panel (d) examines the role of the tilt factor. A higher tilt factor improves the Sharpe ratio, indicating that giving more emphasis to the LLM score in the final allocation enhances risk-adjusted returns. This finding is consistent with the notion that the LLM extracts meaningful return signals from stock-specific news when guided by a basic prompt directive. Especially, while the improvement appears modest (0.95 to 1.1), the tilt factor's impact is mainly capped by the weight constraints imposed in the portfolio setup. Importantly, this highlights a practical use case: by systematically tilting allocations toward the model’s signals, the contribution of news-driven insights can be amplified without abandoning traditional portfolio construction principles. This creates an attractive balance — maintaining the stability of a benchmark-anchored allocation while still extracting measurable value from alternative data.

    Panel (e) considers the number of stocks in the portfolio. The LLM-enhanced strategy achieves its strongest performance when concentrated in 25 names, with a Sharpe ratio of roughly 1.3, and performance declines as the portfolio broadens to 100 holdings. This suggests that the model’s signal is most effective for a relatively small set of stocks, likely those with clear and impactful news flow. The result also connects to the role of the tilt factor: with fewer holdings, the tilt toward LLM scores can exert a stronger influence on portfolio weights, amplifying the benefits of the signal.
    Such a feature can be especially valuable in contexts where exogenous constraints limit the number of positions that can be held. For a portfolio manager, a less diversified portfolio can help avoid excessive similarity to the benchmark, thereby preserving the potential for meaningful active returns. For a financial advisor, it enables the delivery of a concise, high-conviction list of recommendations to clients, ensuring that the proposed investment set remains selective, focused, and actionable.
    
    Notably, the Sharpe ratio also improves substantially when the portfolio is formed without any pre-filtering of the initial 100 stocks. This is important because such a design removes any selection effect prior screening, leaving performance improvements driven solely by the news-based scores and the tilt factor. In practical terms, this demonstrates that the LLM signal has sufficient standalone power to add value in portfolio construction, even without auxiliary selection rules.

    Finally, Panel (f) examines the impact of different baseline weighting strategies. The results show that performance improves most strongly under a value-weighted allocation when moving from the baseline to the LLM-enhanced strategy. This makes intuitive sense: by focusing more weight on larger firms, which dominate value-weighted portfolios, the model is effectively given greater influence where its informational edge is strongest. Larger companies typically generate more news flow and market commentary, providing the LLM with a richer foundation from which to extract signals. This finding is highly relevant in practice, as investors usually benchmark against market-cap-weighted indices rather than equal-weighted ones. The ability of the LLM-enhanced strategy to deliver excess performance on top of such a realistic baseline not only improves comparability to widely used passive benchmarks but also demonstrates that concentrating the model’s power in news-rich stocks can unlock tangible improvements in risk-adjusted returns.

    Overall, the parameter analysis highlights the robustness of the LLM enhancement across a range of plausible configurations. The strongest gains arise when portfolios are more concentrated, rebalanced less frequently, and when the prompt structure is simple yet combined with a meaningful tilt toward the model’s signals. This underscores that the LLM’s predictive value is best unlocked not through complexity or excessive data, but through a disciplined framework that amplifies the model’s insights within conventional portfolio design.

\section{Conclusion}

    This paper examines whether large language models can enhance a standard cross-sectional momentum strategy by incorporating real-time insights from firm-specific news. Using S\&P 500 constituents, daily equity returns, and high-frequency news data, we implement prompt-engineered interactions with ChatGPT 4.0 mini that explicitly inform the model when a stock is about to enter a momentum portfolio. The LLM then evaluates whether recent news supports a continuation of past returns, producing a score that conditions both stock selection and portfolio weights.

    Our empirical analysis shows that the LLM-enhanced momentum strategy consistently outperforms its baseline counterpart in both in-sample and out-of-sample settings, achieving higher Sharpe and Sortino ratios, lower volatility, and smaller drawdowns, all while maintaining comparable turnover. The improvements are robust to conservative transaction cost assumptions, alternative prompt designs, and portfolio construction constraints. Importantly, the out-of-sample period begins after the model's pre-training cut-off, ensuring that the observed gains cannot be attributed to information leakage.

    These results contribute to the literature on the sources of alpha in momentum strategies, suggesting that part of the return premium arises from delayed incorporation of firm-specific information into prices. By processing news in real time, LLMs can extract signals that complement traditional return-based predictors and help exploit these informational frictions more effectively.

    Our study has limitations. The out-of-sample period is relatively short, which limits statistical power, and the estimated alpha from time-series regressions is only weakly significant in some cases. The analysis is restricted to a single pre-trained model and a large-cap U.S. equity universe, leaving open questions about generalizability to other markets, asset classes, or less liquid securities.

    Future research could extend this framework by exploring fine-tuning or domain adaptation of LLMs to financial text, integrating signals from alternative data sources, or testing in multi-factor portfolio settings. As more data become available, evaluating long-horizon performance will be essential for assessing the stability and scalability of LLM-enhanced strategies. Overall, our findings suggest that LLMs are not merely experimental tools but practical, scalable components of modern investment processes, capable of delivering incremental value beyond established factor models.

\newpage
\bibliographystyle{apalike}
\bibliography{biblio}

\appendix

\end{document}